\documentclass[twocolumn]{aastex62}

\newcommand{\Mjup}{\mbox{$M_\mathrm{Jup}$}}
\newcommand{\Msun}{\mbox{$M_{\odot}$}}

\shorttitle{A Substellar Companion Orbiting HD 47127}
\shortauthors{Bowler et al.}

\begin{document}

\title{The McDonald Accelerating Stars Survey (MASS): \\
Discovery of a Long-Period Substellar Companion Orbiting the Old Solar Analog HD 47127}

\correspondingauthor{Brendan P. Bowler}
\email{bpbowler@astro.as.utexas.edu}

\author[0000-0003-2649-2288]{Brendan P. Bowler}
\affiliation{Department of Astronomy, The University of Texas at Austin, Austin, TX 78712, USA}

\author{Michael Endl}
\affiliation{McDonald Observatory and the Department of Astronomy, The University of Texas at Austin, Austin, TX 78712, USA}

\author{William D. Cochran}
\affiliation{Center for Planetary Systems Habitability and McDonald Observatory, The University of Texas at Austin, Austin, TX 78712, USA}

\author{Phillip J. MacQueen}
\affiliation{McDonald Observatory and the Department of Astronomy, The University of Texas at Austin, Austin, TX 78712, USA}

\author{Justin R. Crepp}
\affiliation{Department of Physics, University of Notre Dame, 225 Nieuwland Science Hall, Notre Dame, IN 46556, USA}

\author{Greg W. Doppmann}
\affiliation{W. M. Keck Observatory, 65-1120 Mamalahoa Hwy., Kamuela, HI 96743, USA}

\author{Shannon Dulz}
\affiliation{Department of Physics, University of Notre Dame, 225 Nieuwland Science Hall, Notre Dame, IN 46556, USA}

\author{Timothy D. Brandt}
\affiliation{Department of Physics, University of California, Santa Barbara, Santa Barbara, CA 93106, USA}

\author[0000-0003-0168-3010]{G.~Mirek Brandt}
\altaffiliation{NSF Graduate Research Fellow}
\affiliation{Department of Physics, University of California, Santa Barbara, Santa Barbara, CA 93106, USA}

\author{Yiting Li}
\affiliation{Department of Physics, University of California, Santa Barbara, Santa Barbara, CA 93106, USA}

\author[0000-0001-9823-1445]{Trent J.~Dupuy}
\affiliation{Institute for Astronomy, University of Edinburgh, Royal Observatory, Blackford Hill, Edinburgh, EH9 3HJ, UK}

\author[0000-0003-4557-414X]{Kyle Franson}
\affiliation{Department of Astronomy, The University of Texas at Austin, Austin, TX 78712, USA}

\author{Kaitlin M. Kratter}
\affiliation{Department of Astronomy, University of Arizona, Tucson, AZ 85721, USA}

\author{Caroline V. Morley}
\affiliation{Department of Astronomy, The University of Texas at Austin, Austin, TX 78712, USA}

\author{Yifan Zhou}
\affiliation{Department of Astronomy, The University of Texas at Austin, Austin, TX 78712, USA}



\begin{abstract}
Brown dwarfs with well-determined ages, luminosities, and masses provide rare but 
valuable tests of low-temperature atmospheric and evolutionary models.
We present the discovery and dynamical mass measurement of a substellar companion to HD 47127, 
an old ($\approx$7--10~Gyr) G5 main sequence star with a mass similar to the Sun.
Radial velocities of the host star with the Harlan J. Smith Telescope uncovered a low-amplitude  
acceleration of 1.93~$\pm$~0.08 m s$^{-1}$ yr$^{-1}$
based on 20 years of monitoring.
We subsequently recovered a faint ($\Delta H$=13.14 $\pm$ 0.15~mag) 
co-moving companion at 1.95$''$ (52 AU)  
with follow-up Keck/NIRC2 adaptive optics imaging.
The radial acceleration of HD 47127 together with its tangential  
acceleration from \emph{Hipparcos} and \emph{Gaia} EDR3 astrometry provide a
direct measurement of the three-dimensional acceleration vector of the host star, enabling a
dynamical mass constraint for HD 47127 B (67.5--177~\Mjup \ at 95\% confidence) 
despite the small fractional
orbital coverage of the observations.
The absolute $H$-band magnitude of HD 47127 B is fainter than
the benchmark T dwarfs HD~19467~B and Gl~229~B but brighter than Gl~758~B and HD~4113~C,
suggesting a late-T spectral type.  
Altogether the mass limits for HD 47127 B from its dynamical mass and the substellar boundary
imply a range of 67--78~\Mjup \ assuming it is single, although a preference for high masses
of $\approx$100~\Mjup \ from  dynamical constraints
hints at the possibility that HD 47127 B could itself be a binary pair of brown dwarfs 
or that another massive companion resides closer in.
Regardless, HD 47127 B will be an excellent target for more refined orbital and atmospheric characterization in the future. \\

\end{abstract}


\section{Introduction} \label{sec:intro}

Brown dwarfs and giant planets lack sufficient core temperatures and pressures to stably fuse hydrogen,   
forcing them to inexorably cool and grow dimmer over time (\citealt{Kumar:1963ht}).  The pace of their evolution is predominantly dictated by their mass but also their accretion history, initial entropy, metallicity, 
and atmospheric opacity, which varies with wavelength and effective temperature 
(e.g., \citealt{Marley:2007bf}; \citealt{Baraffe:2009jw}; \citealt{Spiegel:2012ea}; \citealt{Marleau:2014bh}).  
As they cool, substellar objects pass through the L,
T, and Y spectral classes, which are defined based on the strength of 
spectroscopic absorption features in the optical and near-infrared
such as FeH, TiO, VO, H$_2$O, CO, CH$_4$, and NH$_3$ 
(e.g., \citealt{Kirkpatrick:1999ev}; \citealt{Burgasser:2006cf}; \citealt{Cushing:2011dk}).
Physically, these features 
are tied to complex time-dependent chemical and physical 
processes including grain formation, rain out of condensates,
vertical mixing, and 
chemical disequilibrium (e.g., \citealt{Ackerman:2001gk}; \citealt{Helling:2008gs}; \citealt{Morley:2012io}).
These atmospheric phenomena in turn affect thermal evolution 
by acting as surface boundary conditions for interior structure models. 
Because so many factors can influence substellar evolution, there is a critical need to
empirically validate low-temperature cooling models across a wide range of masses and ages.
This is especially urgent because of their regular use to infer the masses of 
directly imaged exoplanets (\citealt{Bowler:2016jk}).

Brown dwarfs with dynamically measured masses are among the most valuable tools to
test evolutionary models.
One approach is to patiently monitor the orbits of brown dwarf binaries; for example, 
the first binary T dwarf with a dynamical mass revealed early discrepancies
between the properties inferred from atmospheric models and those of evolutionary models (\citealt{Liu:2008ib}).
Similarly, \citet{Konopacky:2010kr} found that widely used evolutionary models  both
systematically under- and over-predict the masses of ultracool dwarfs, and
\citet{Dupuy:2017ke} found that ``hybrid'' evolutionary models developed by \citet{Saumon:2008im} 
in which clouds dissipate 
at the L/T transition are most consistent with coevality tests.
When masses \emph{and} ages are both independently constrained, 
as in the case of brown dwarf companions to stars, evolutionary models
can be directly tested 
(e.g., \citealt{Dupuy:2009wd}; \citealt{Dupuy:2014iz}; \citealt{Bowler:2018gy}; \citealt{Brandt:2019ey}).
About 15 dynamical masses of substellar companions have been measured,
only six of which are for T dwarfs.

Brown dwarf companions with well-determined orbits are also valuable to
probe the formation route of these objects as a population.
\citet{Bowler:2020hk} found that the orbits of directly imaged giant planets between 5--100~AU are
significantly more circular compared to brown dwarf companions, which peak at 
eccentricities of $\approx$0.6--0.9 and more 
closely resemble the orbital properties of wide stellar binaries.  However, these results
were based on a  sample of 27 objects (9 giant planets and 18 brown dwarfs);
more examples are needed to address nuanced questions about 
how eccentricities might vary with age and stellar host mass.
Fortunately, several recent and ongoing programs are using stellar accelerations from long-baseline
radial velocity (RV) surveys and astrometry from \emph{Hipparcos} and \emph{Gaia} to
identify new benchmark brown dwarf companions whose orbits can be well constrained 
(e.g., \citealt{Crepp:2014ce}; \citealt{Rickman:2020aa}; \citealt{Maire:2020iu}; \citealt{Currie:2020hq}).

Here we present the direct imaging discovery of a substellar companion to HD 47127,
a Sun-like G5 main sequence star (\citealt{Adams:1935aa}) 
located at a distance of 26.62 $\pm$ 0.02 pc 
(based on $Gaia$ EDR3; \citealt{GaiaCollaboration:2020ev}).
HD 47127 is slightly metal-rich ([Fe/H] = +0.1 dex; \citealt{Valenti:2005fz})
with a mass of 1.02 $\pm$ 0.05~\Msun \ (\citealt{Luck:2017jd}).
It is also inactive and old: \citet{Isaacson:2010gka} find a log~$R'_{HK}$ value of --4.984 dex
and \citet{Wright:2004eb} measure a value of --5.02 dex; both imply an age of $\approx$6.3--7.0~Gyr
using empirically calibrated age-activity relations from \citet{Mamajek:2008jz}
and an age of 8.7$\pm$3.2 Gyr (with a 95\% CI between 2.7--13~Gyr) 
using \texttt{BAFFLES}, a Bayesian-based age-dating tool (\citealt{StanfordMoore:2020kw}).
This is in good agreement with isochronal ages of 9.0$^{+3.8}_{-1.7}$ Gyr from
the Geneva-Copenagen Survey of the Solar Neighborhood (\citealt{Nordstrom:2004ci}) and
7.8$^{+3.0}_{-3.7}$ Gyr from \citet{Valenti:2005fz}.
The $v \sin i$ value for HD 47127 is 1.8 km s$^{-1}$ (\citealt{Glebocki:2005aa}), 
consistent with it being 
a slow rotator with a long rotation period.

The first sign that HD 47127 harbors a wide-separation companion 
emerged from a shallow RV trend obtained with 
the Tull Spectrograph at the Harlan J. Smith telescope during the 
McDonald Observatory Planet Search Program (MOPS; \citealt{Cochran:1993va}).
We subsequently recovered a faint comoving companion
with adaptive optics imaging using NIRC2 at Keck Observatory
as part of the McDonald Accelerating Stars Survey (MASS), a follow-up 
high-contrast imaging campaign to identify the nature of
these long-term radial accelerations (\citealt{Bowler:2018gy}; \citealt{Bowler:2021gg}). 
HD 47127 B is located at a projected separation of 1$\farcs$95 (52 AU)
and orbital motion is evident between 2017 to 2020. 
Based on its absolute $H$-band magnitude, HD 47127 B is expected to be a late-T dwarf.
In Section~\ref{sec:obs} we describe our RV and imaging observations of HD 47127.
Astrometric measurements, demonstration of common proper motion,
and results of the orbit fit can be found in Section~\ref{sec:results}.
Finally, HD 47127 B is discussed in the broader context of other benchmark substellar companions
in Section~\ref{sec:discussion}.

\section{Observations} \label{sec:obs}

\subsection{Radial Velocities} \label{sec:tull}

\begin{figure}
  \vskip -0.2 in
  \hskip -0.8 in
  \resizebox{5in}{!}{\includegraphics{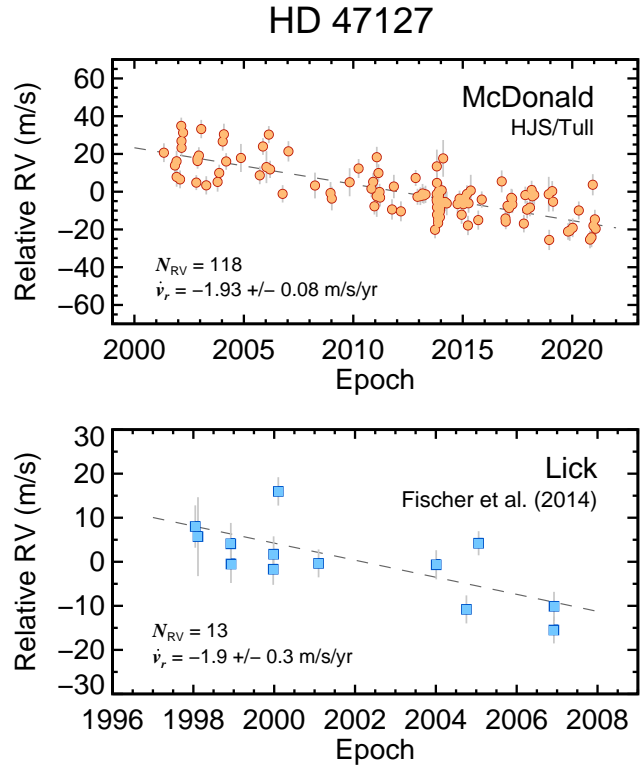}}
  \vskip -2.5 in
  \caption{Radial velocities from McDonald Observatory (top) and 
  Lick Observatory (bottom).  Both datasets show a shallow radial acceleration
  of $\approx$--1.9 m s$^{-1}$ yr$^{-1}$.   \label{fig:hd47127_rvs} } 
\end{figure}

A total of 118 high-resolution ($R$ $\equiv$ $\lambda$/$\delta \lambda$ $\approx$ 60,000) 
spectra were acquired between 2001 and 2021 with the
Tull Coud\'{e} spectrograph (\citealt{Tull:1995tn}) using the 1$\farcs$2 slit at McDonald Observatory's 
2.7-m Harlan J. Smith telescope.
A molecular iodine cell was mounted in the light path to measure RVs with
respect to an iodine-free template following the procedure in \citet{Endl:2000ui}.
The very small secular acceleration for HD~47127 was removed.
Results are shown in Figure~\ref{fig:hd47127_rvs} and RVs are listed in Table~\ref{tab:hd47127_rvs}.  A clear linear
trend is evident with a slope of --1.93 $\pm$ 0.08 m s$^{-1}$ yr$^{-1}$.
After subtracting the linear fit, the RMS of the residuals 
is 8.2 m s$^{-1}$, which is about twice the median RV uncertainty
of 4.7 m s$^{-1}$.  This suggests that HD 47127 is slightly active or
that it could harbor an additional inner companion.
However, its Mount Wilson Observatory $S$-index is low; we find a value of 0.158 $\pm$ 0.012 from our Tull spectra, 
which was derived by first computing the McDonald $S$-index 
and then transforming it to the Mount Wilson system following \citet{Paulson:2002tn}.
This is consistent with other activity measurements for HD 47127 (\citealt{Isaacson:2010gka}; \citealt{Wright:2004eb}) 
and is similar to the Sun at minimum activity as well as old field stars of the same spectral type (e.g., \citealt{Saikia:2018dh}).
A Lomb-Scargle periodogram of the RV residuals does not reveal any significant peaks, so
it is unclear if the excess RV scatter originates from one or more inner planets or perhaps from a modest level of stellar activity.

\begin{deluxetable}{lcc}
\renewcommand\arraystretch{0.9}
\tabletypesize{\small}
\setlength{ \tabcolsep } {.1cm}
\tablewidth{0pt}
\tablecolumns{3}
\tablecaption{Tull Spectrograph Relative Radial Velocities of HD47127\label{tab:hd47127_rvs}}
\tablehead{
    \colhead{Date}  & \colhead{RV} & \colhead{$\sigma_{\mathrm{RV}}$} \\
    \colhead{(BJD)} & \colhead{(m s$^{-1}$)} & \colhead{(m s$^{-1}$)}
        }
\startdata
   2452037.59844    &        20.7    &         5.0   \\
   2452221.93066    &        13.9    &         3.8   \\
   2452248.87150    &        16.1    &         5.6   \\
   2452249.85430    &         7.8    &         5.2   \\
   2452306.76529    &         6.5    &         4.7   \\
   2452326.70915    &        34.9    &         4.3   \\
   2452330.68521    &        26.8    &         4.3   \\
   2452331.66394    &        23.3    &         4.2   \\
   2452357.65539    &        31.2    &         4.5   \\
   2452577.97230    &         4.8    &         4.0   \\
\multicolumn{3}{c}{$\cdots$} \\
\enddata
\tablecomments{Table 1 is published in its entirety in the machine-readable format.
      A portion is shown here for guidance regarding its form and content.}
\end{deluxetable}

HD 47127 was also targeted with the Hamilton Spectrograph as part of the 
Lick Planet Search (\citealt{Fischer:2014ew}).
Altogether 13 RVs were obtained between 1998 and 2007 with a median velocity
precision of 3.3 m s$^{-1}$ (Figure~\ref{fig:hd47127_rvs}).
We find a slope of --1.9 $\pm$ 0.3 m s$^{-1}$ yr$^{-1}$, which is similar to (but much less precise than)
the trend from the McDonald RVs.
The RMS level of the residuals is 5.9 m s$^{-1}$.


\subsection{Adaptive Optics Imaging} \label{sec:nirc2}

We obtained natural guide star adaptive optics images of HD 47127 with 
Keck/NIRC2 in its narrow camera mode 
on the nights of 2017 October 10 UT, 2020 December 31 UT,
and 2021 January 22 UT.
HD 47127 was centered behind the 600 mas diameter partly transparent coronagraph mask 
for all of the observations. 
Dome flats were taken at the start of each night.

The October 2017 data consists of a single $H$-band frame with an integration time of 10~seconds and 
one coadd taken at the end of the night as the sky was brightening.
Conditions were photometric with excellent seeing (0$\farcs$4--0$\farcs$5) throughout the night.
Two point sources were visible---one brighter object at $\approx$5$\farcs$6 from HD 47127 
(a background star) and a faint object embedded in the speckle pattern at $\approx$2$''$ (HD 47127 B).
The December 2020 dataset consists of 11 $H$-band images
each with a single coadd taken in ``vertical angle'' (pupil-tracking) mode.  
The total field of view rotation was 6.1$\degr$ for this angular differential
imaging (ADI; \citealt{Liu:2004kk}; \citealt{Marois:2006df}) sequence.  
The January 2021 observations consist of an ADI sequence in $K_S$ band 
taken over the course of about 45 minutes.
Each image has an integration time of 2 seconds per coadd with 10 coadds,
resulting in a total exposure time of 20~s per frame.
Cloud cover and seeing were highly variable; out of 80 frames, 68 were retained
resulting in a field of view rotation of 59$\degr$.

\begin{figure*}
  \vskip -0.8 in
  \resizebox{7.5in}{!}{\includegraphics{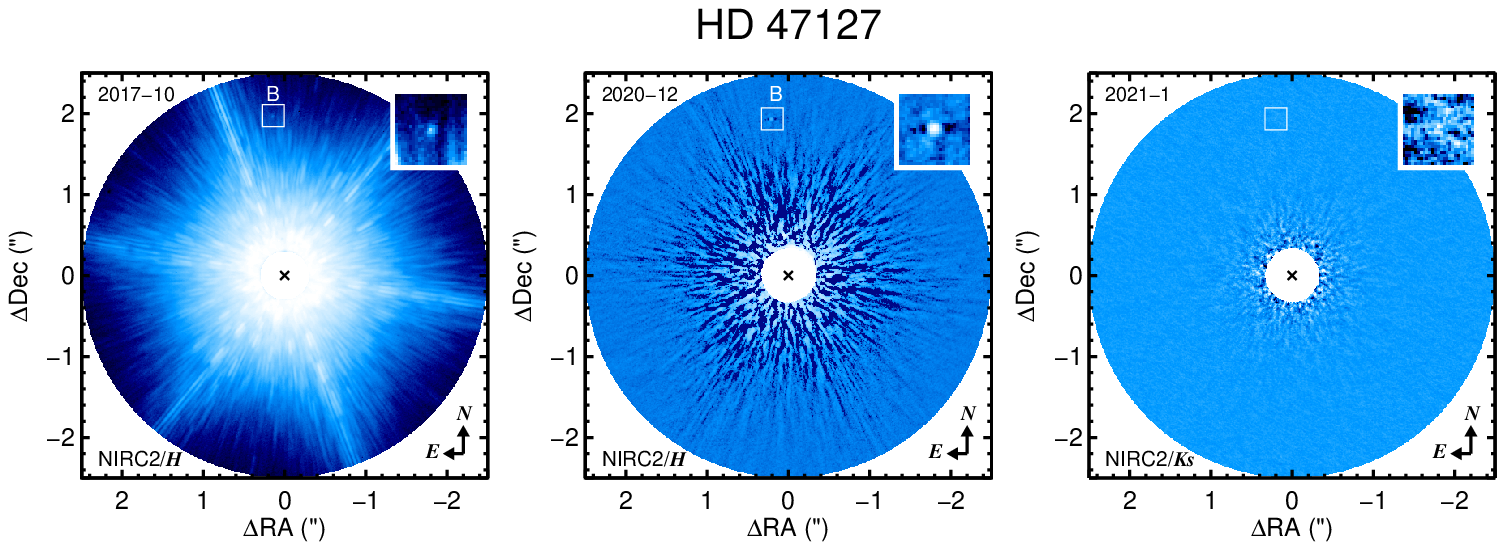}}
  \vskip -7.3 in
  \gridline{\fig{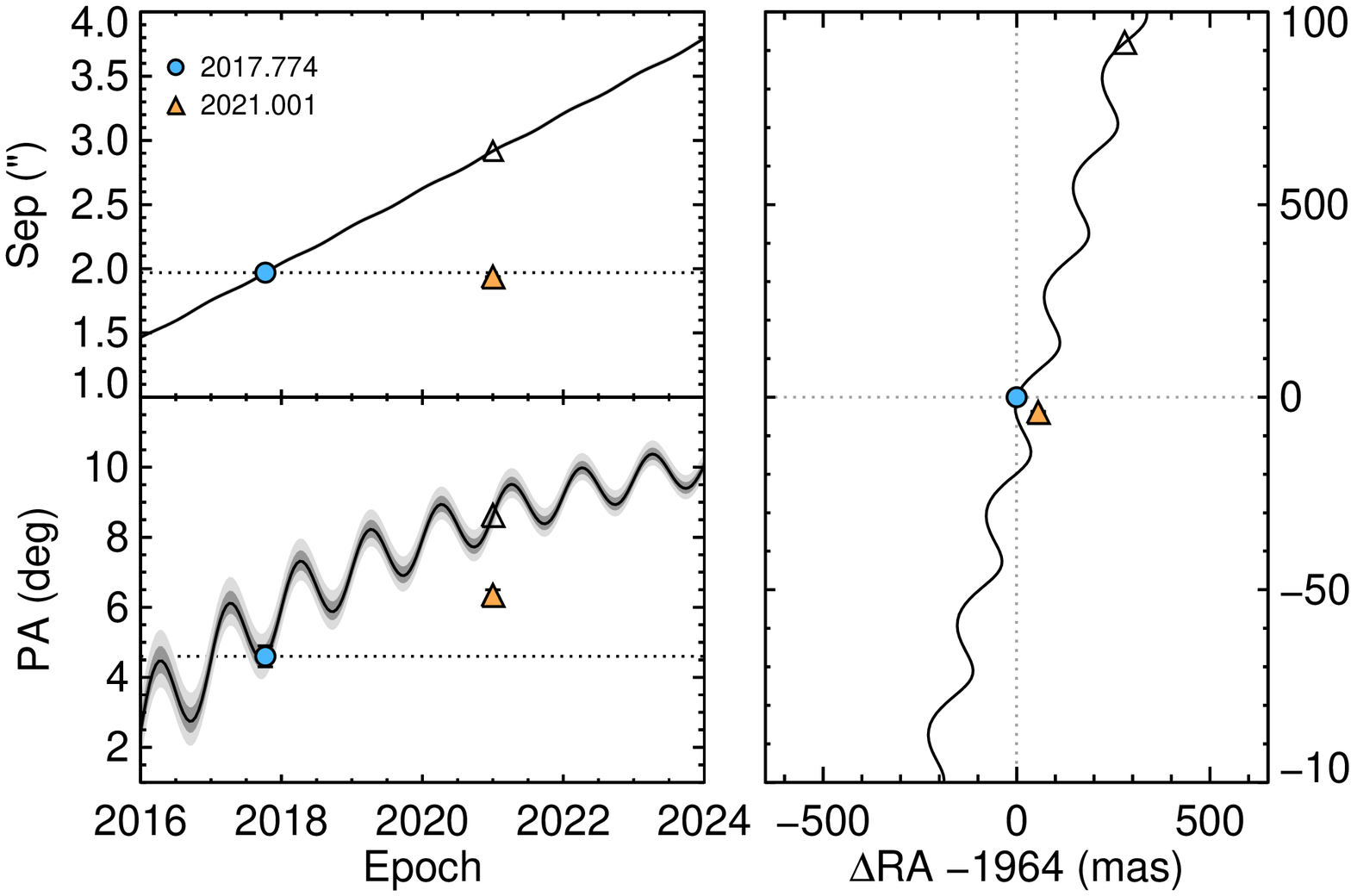}{0.5\textwidth}{}
                \fig{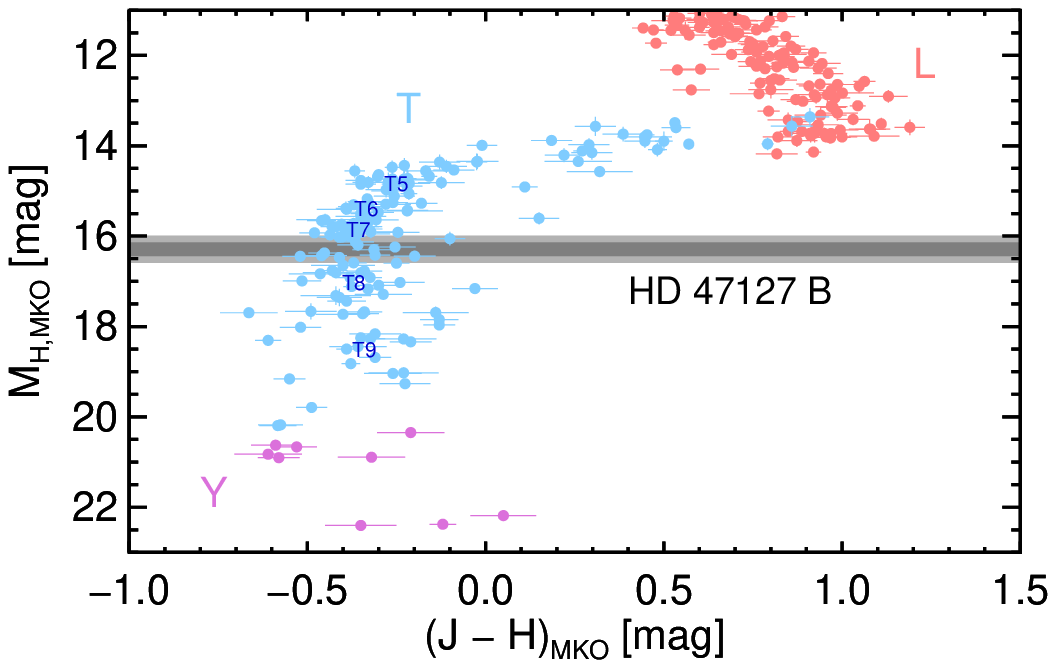}{0.5\textwidth}{}}               
  \vskip -.3 in                
  \caption{\emph{Upper panels:} Adaptive optics images of HD 47127 with Keck/NIRC2.  In each observation the star (denoted with an ``$\times$'') is positioned behind the 600 mas-diameter coronagraph.  HD 47127 B is visible in a single raw $H$-band frame in October 2017 (upper left) and the PSF-subtracted $H$-band sequence in December 2020 (upper middle).  We do not detect the companion in a $K_S$-band ADI sequence in January 2021 (upper right).  The insets show zoomed-in views centered on the companion or its expected location, as in the case of the non-detection in 2021.  North is up and east is to the left.  \emph{Bottom left panels:} Expected trajectory of a background star (solid curve) relative to the initial astrometry in October 2017 (filled circle).  Our second epoch in December 2020 (filled triangle) clearly establishes that HD 47127 B is bound with signs of orbital motion, especially in P.A.  The open triangle shows its expected position if it was a background star.  \emph{Bottom right panel:} Near-infrared color-magnitude diagram showing our December 2020 absolute $H$-band magnitude measurement of HD 47127 B compared to L, T, and Y dwarfs from \href{bit.ly/UltracoolSheet}{\emph{The UltracoolSheet}}.  The gray shaded bands represent the 1$\sigma$ and 2$\sigma$ uncertainties.  Based on this constraint, we expect HD 47127 B to be a late-T dwarf ($\approx$T5--T8). \label{fig:nirc2} } 
\end{figure*}

Image reduction and PSF subtraction of the December 2020 and January 2021 
ADI sequences follows the description 
in \citet{Bowler:2015ja}.  Raw frames are cleaned of cosmic
rays and bad pixels, bias subtracted, and flat fielded.  Images are then registered using the measured position of the
host star behind the coronagraph and PSF subtraction is carried out using the
Locally Optimized Combination of Images algorithm (\citealt{Lafreniere:2007bg}).
The final processed frames are shown in Figure \ref{fig:nirc2}.
A nearby point source is visible in the single $H$-band image from October 2017
and the processed $H$-band frame from December 2020, but it is not present
in the $K_S$-band data from January 2021.


\begin{figure*}
  \vskip -0.1 in
  \hskip 0.5 in
  \gridline{\fig{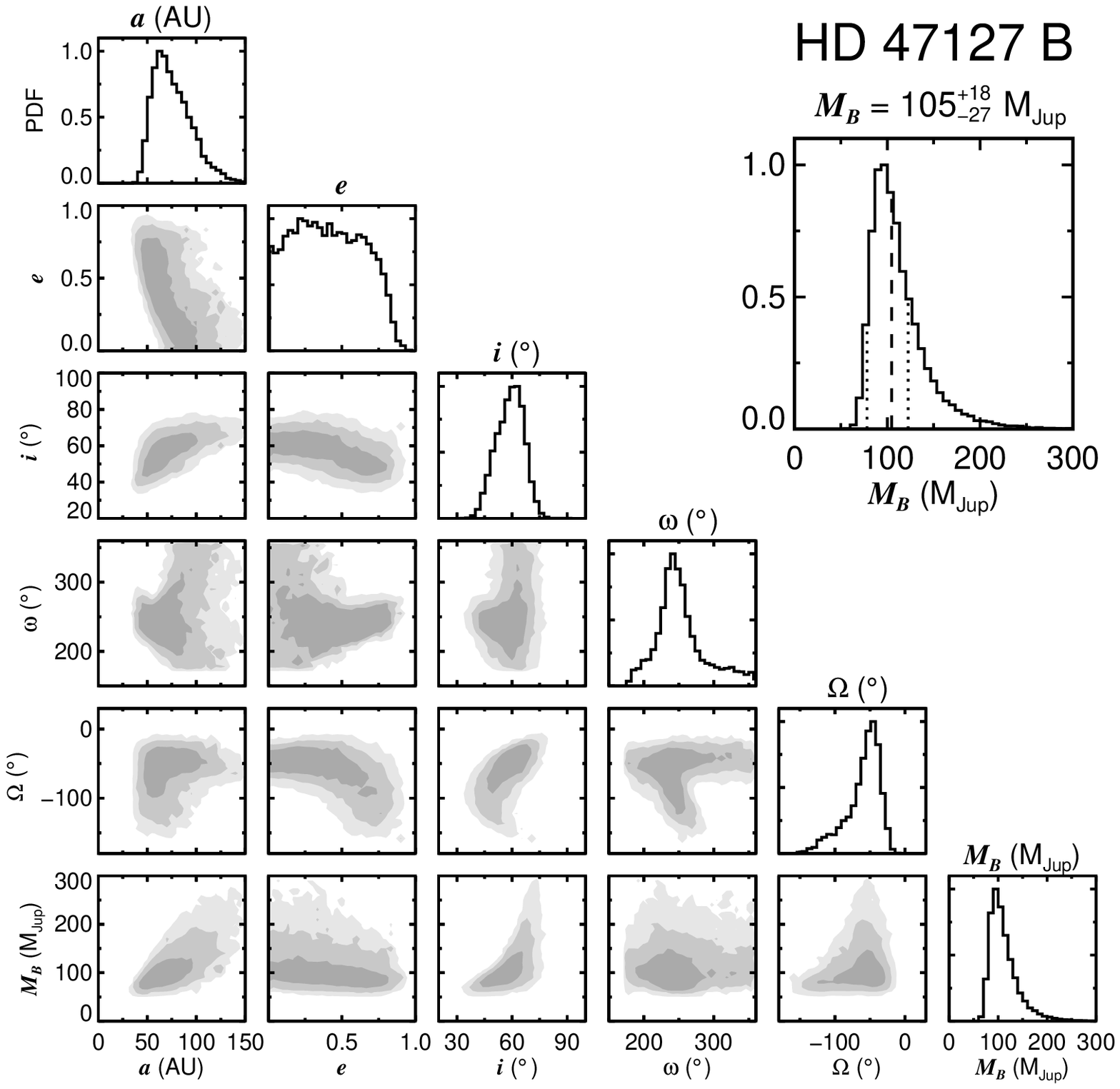}{0.85\textwidth}{}}               
  \vskip -.75 in                
  \gridline{\fig{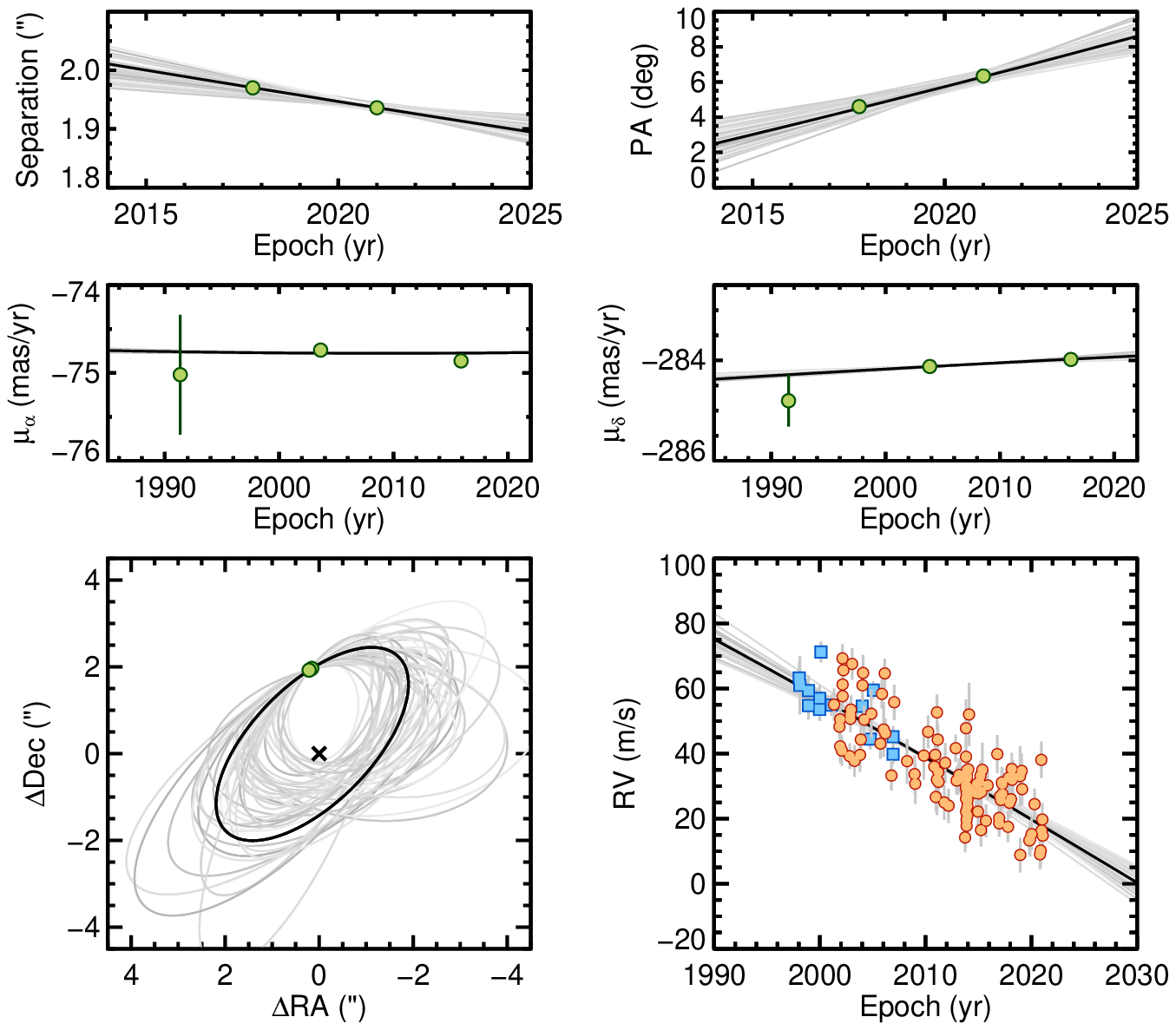}{0.75\textwidth}{}}               
  \vskip -3.35 in                
  \caption{Results of the orbit fit for HD 47127 B.  \emph{Top:} Corner plot showing the joint distributions of orbital elements and their marginalized distributions.  The inset highlights the companion mass distribution, which is broad and allows for both stellar and substellar masses.  If the true mass is near the median value of 105~\Mjup, this would indicate the companion is itself a binary T dwarf or that an additional inner companion is biasing the observed acceleration.  \emph{Bottom Panels:} Randomly drawn orbits from the posterior MCMC chains compared to the relative astrometry; HGCA proper motions from \emph{Hipparcos}, \emph{Gaia}, and the scaled positional difference between the two missions; and radial velocities.  \label{fig:orbit} } 
\end{figure*}

\section{Results} \label{sec:results}

\subsection{Relative Astrometry and Photometry} \label{sec:astrometry}

\subsubsection{October 2017 Observations}

Aperture photometry is used to calculate the flux ratio of HD 47127 B
for the single coronagraphic frame from October 2017.
A ``raw'' flux value comprising the source, the sky background, and the local PSF wing of
HD 47127 is calculated using a circular aperture radius of 4 pix (one diffraction width at 1.6~$\mu$m)
centered on the centroid 
position of the companion.  To estimate the background flux at that location,
48 non-overlapping identical apertures are sampled at the same separation
as HD 47127 B but distributed azimuthally around the host star while avoiding
the six diffraction spikes as well as HD 47127 B.
The mean and standard deviation of these apertures are adopted
as the local background level (sky plus stellar PSF) to 
sample variations in the speckle pattern in the image,
and the true flux of HD 47127 B is taken to be the raw flux minus
the mean of these local background values.
The flux of HD 47127 A ($f_A$) is determined using the raw
aperture-summed flux of the host star ($f_{r,A}$), the background sky level
of the image ($s$), and the coronagraph 
throughput ($t_H$) in $H$ band of 0.099$\pm$0.013\% (7.51 $\pm$ 0.14 mag) from \citet{Bowler:2015ja} 
as follows: $f_A$ = ($f_{r,A} + s)/t_H - s$.
This applies a small correction to the primary flux to take into account 
the fact that the coronagraph is attenuating the star-plus-sky flux,
not just that of the host star.
We measure a contrast of $\Delta H$ = 12.5 $\pm$ 0.6~mag for HD 47127 B, 
which includes uncertainties in the coronagraph throughput and companion background
levels.

Astrometry is computed using the centroid positions of the primary and companion.
To estimate the positional uncertainty, we inject a 2D Gaussian with the same amplitude 
and standard deviation as HD 47127 B at each of the 48 azimuthally distributed positions around the host
star mentioned above.  We then attempt to recover the locations of each of the synthetic companions
using the same centroid positional approach we used for HD 47127 B.
The maximum of these measurements is 0.7~pix, which we adopt as a conservative estimate of the uncertainty
in our centroid measurements of the companion.
Our final astrometry for the October 2017 dataset (epoch 2017.774)
is $\rho$ = 1.970 $\pm$ 0.010$''$ and $\theta$=4.6 $\pm$ 0.3$\degr$,
which takes into account uncertainties due to random
measurement errors, plate scale precision, distortion solution, and true north orientation
(for the P.A. value) following \citet{Bowler:2015ja}.

\subsubsection{December 2020 Observations}

Astrometry and relative photometry for the December 2020 ADI sequence
is determined using the negative companion injection approach described
in \citet{Bowler:2018gy}.  We did not obtain unsaturated frames to independently
flux calibrate the deep sequence, so for a PSF model 
we adopt the median-combined $H$-band unsaturated image of the bright star 12 Psc 
from \citet{Bowler:2021gg} taken on 2017 October 10 UT.
The PSF model flux is normalized to the mean of the peak values of HD 47127
behind the mask.
The PSF model amplitude, separation, and P.A. are then iteratively adjusted using the \texttt{amoeba}
algorithm (\citealt{Nelder:1965tk}) to remove the signal of the companion in the processed image and
minimize the RMS in a circular aperture at the position of the companion.  
After taking into account the mask transmission and associated uncertainty, 
this results in a contrast of $\Delta H$ = 13.14~$\pm$~0.15~mag---in agreement
with our measurement from October 2017 at the 1-$\sigma$ level.
Our best fit astrometry for the December 2020 dataset (epoch 2020.999)
is $\rho$ = 1.936 $\pm$ 0.003$''$ and $\theta$=6.34 $\pm$ 0.16$\degr$,
which also takes into account PSF blurring effects caused by field of view
rotation within each exposure.

\subsubsection{January 2021 Observations}

HD 47127 B is not detected in our processed January 2021 dataset, but a non-detection
in $K_S$ band nevertheless constrains the $H$--$K_S$ color of the companion.
A lower limit on the $K_S$-band contrast is determined using aperture
photometry of the primary star and the RMS counts at the location of the companion, 
which was determined from our December 2020 astrometry taken only a few weeks earlier.
The average flux of the primary in a circular 4-pix radius aperture from all 68 frames 
in the ADI sequence is adopted for the host.  A correction is applied to take into
account the 0.22 $\pm$ 0.02\% throughput of the 600~mas coronagraph mask
from \citet{Bowler:2015ja}.  Using a 3-$\sigma$ flux upper limit for the companion,
we find a contrast of $\Delta K_S$ $>$ 11.6~mag.  
This implies that HD~47127 has an $H$--$K_S$ color of $<$1.7~mag based
on the 2MASS magnitudes for the host. 

\subsection{Common Proper Motion} \label{sec:cpm}

Figure~\ref{fig:nirc2} shows the expected motion of a background star
relative to the initial astrometry from 2017 based on the
proper motion and parallax of the host star from $Gaia$ EDR3 
($\mu_{\alpha}  \cos \delta$ = --74.864 $\pm$ 0.028 mas yr$^{-1}$, 
$\mu_{\delta}$ = --283.982 $\pm$ 0.021 mas yr$^{-1}$, 
$\pi$ = 37.561 $\pm$ 0.025; \citealt{GaiaCollaboration:2020ev}).
A stationary background star should substantially increase in both separation
and P.A. over time.  HD 47127 B is clearly comoving with its
host star and shows slight orbital motion, mostly in P.A.
A linear fit to the orbital motion gives $\rho (t)$ = --10.5 $\pm$ 3.2 mas yr$^{-1}$
for the change in separation and $\theta (t)$ = 0.54 $\pm$ 0.11$\degr$ yr$^{-1}$
for P.A.

\subsection{Orbit Fit and Dynamical Mass}{\label{sec:orbit}}

Astrometry of HD 47127 was obtained by both \emph{Hipparcos}
and \emph{Gaia}, but does not show a significant acceleration
in \citet{Brandt:2018dja}'s \emph{Hipparcos}-\emph{Gaia} Catalog of Accelerations (HGCA).
That catalog was based on \emph{Gaia} DR2 and gives an
acceleration of $d \mu_{\alpha \delta}/dt$ =1.6 $\pm$ 1.1 m s$^{-1}$ yr$^{-1}$, or an SNR of 1.5,
using the \emph{Gaia} and \emph{Hipparcos}-\emph{Gaia} 
scaled positional difference measurements following \citet{Brandt:2019ey}.  
More recently, Brandt et al. (submitted) updated this catalog with
\emph{Gaia} EDR3.\footnote{In the \emph{Gaia} EDR3 catalog, HD 47127 is listed as having 
RUWE=1.131, $\chi^2$=711 (for 192 good AL observations), and astrometric excess noise=0.14 mas.  To assess whether these
values are unusual for stars like HD 47127, we queried the \emph{Hipparcos} catalog for bright G2--G5 main sequence
stars with $V$=6.3--7.3 (within $\pm$0.5 mag of HD 47127).  This resulted in 398 stars,
which were then cross-matched in \emph{Gaia}.
The median RUWE value, $\chi^2$ value, and excess noise parameters are 1.03, 1032 (for a median of 316 good AL observations), 
and 0.129 mas, respectively.  
This suggests that the astrometric quality of HD 47127 in \emph{Gaia} is typical for a star of this brightness.}
The proper motion differences in R.A. and Dec. for HD 47127 are
$\Delta \mu_{\alpha, \mathrm{Gaia-HG}}$ = --0.12 $\pm$ 0.05 mas yr$^{-1}$
and $\Delta \mu_{\delta, \mathrm{Gaia-HG}}$ = 0.13 $\pm$ 0.03 mas yr$^{-1}$, respectively.
With this improved precision, HD 47127 shows a significant
acceleration of 
$d \mu_{\alpha}/dt$ = --1.3 $\pm$ 0.5 m s$^{-1}$ yr$^{-1}$ in R.A.,
$d \mu_{\delta}/dt$ = 1.4 $\pm$ 0.3 m s$^{-1}$ yr$^{-1}$ in Dec.,
and a total tangential acceleration of $d \mu_{\alpha \delta}/dt$ = 1.9 $\pm$ 0.4 m s$^{-1}$ yr$^{-1}$ (SNR=5.0).
This is very similar to the RV trend of 1.93 $\pm$ 0.08 m s$^{-1}$ yr$^{-1}$
from our Tull spectrograph observations.  Together these tangential and radial accelerations
define a 3-dimensional acceleration vector, mapping out the reflex motion of the
host star under the influence of a companion.  Assuming this originates from 
the imaged companion HD 47127 B, this enables an orbit fit and direct dynamical mass measurement of this object.

\begin{deluxetable*}{lcccccc}
\renewcommand\arraystretch{0.9}
\tabletypesize{\small}
\setlength{ \tabcolsep } {.1cm} 
\tablewidth{0pt}
\tablecolumns{7}
\tablecaption{HD 47127 B Orbit Fit Results \label{tab:hd47127b_orbit}}
\tablehead{
    \colhead{Parameter} & \colhead{Prior} & \colhead{Best Fit\tablenotemark{a}} & \colhead{Median}  & \colhead{MAP\tablenotemark{b}}  & \colhead{68.3\% CI}  & \colhead{95.4\% CI}  
        }   
\startdata
\cutinhead{Fitted Parameters}
$M_1$ (M$_{\odot}$) & $\mathcal{N}$(1.02,0.10) & 1.0 & 1.02 & 1.03 & (0.91, 1.12) & (0.82, 1.22) \\
$M_2$ (M$_\mathrm{Jup}$) & 1/$M_2$ & 103 & 105 & 95 & (78, 123) & (68, 177) \\
$a$ (AU) &  1/$a$ & 74 & 73 & 65 & (52, 89) & (44, 118) \\
$\sqrt{e} \sin \omega$ & $\mathcal{U}$(--1,1) & --0.44 & --0.51 & --0.64 & (--0.83, --0.30) & (--0.91, 0.15) \\
$\sqrt{e} \cos \omega$ & $\mathcal{U}$(--1,1) & --0.06 & --0.24 & --0.35 & (--0.54, --0.02) & (--0.61, 0.44) \\
$i$ ($^{\circ}$)& sin$i$ & 62 & 59 & 61 & (52, 68) & (44, 72) \\
$\Omega$ ($^{\circ}$) & $\mathcal{U}$(--180, 180) & --41 & --54 & --49 & (--71, --31) & (--117, --22) \\
$\lambda_{\mathrm{ref}}$ ($^{\circ}$)\tablenotemark{c} & $\mathcal{U}$(--180, 180) & 53 & 82 & 68 & (35, 124) & (17, 205) \\
$\sigma_\mathrm{jit}$ (m/s) & 1/$\sigma_\mathrm{jit}$ & 6.0 & 6.2 & 6.2 & (5.5, 6.8) & (4.9, 7.5) \\
\cutinhead{Derived Parameters}
$e$ &  $\cdots$  &0.20 & 0.40 & 0.33 & (0.12, 0.65) & (0.0, 0.79) \\
$\omega$ ($^{\circ}$)& $\cdots$  & 263 & 247 & 244 & (216, 275) & (180, 334) \\
$P$ (yr) &  $\cdots$  &610 & 590 & 530 & (340, 770) & (270, 1200) \\
$\tau$ (yr)\tablenotemark{d} &  $\cdots$  &2360 & 2240 & 2130 & (2090, 2370) & (2080, 2810) \\
$d_p$ (AU) &  $\cdots$  &59 & 43 & 21 & (12, 66) & (7.3, 100) \\
$d_a$ (AU) &  $\cdots$  &89 & 100 & 92 & (78, 110) & (71, 150) \\
\enddata
\tablenotetext{a}{Maximum likelihood orbit.}
\tablenotetext{b}{Maximum a posteriori probability, or mode of the marginalized posterior distribution.}
\tablenotetext{c}{Mean longitude at the reference epoch, 2455197.5 JD.}
\tablenotetext{d}{Time of periastron, 2455197.5 JD -- $P$($\lambda_{\mathrm{ref}}$ -- $\omega$)/(2$\pi$).}
\end{deluxetable*}

The orbit and dynamical mass of HD 47127 B are constrained using the efficient Bayesian orbit fitting code
\texttt{orvara} (Brandt et al., submitted).  \texttt{orvara} incorporates absolute astrometry
from HGCA, radial velocities (in this case from both McDonald and Lick), and relative astrometry
and fits for nine parameters
using MCMC: the primary mass ($M_1$), the companion mass ($M_2$), the orbital semi-major axis ($a$),
two terms relating the eccentricity ($e$) and  argument of periastron ($\omega$)---$\sqrt{e} \sin \omega$
and  $\sqrt{e} \cos \omega$---the inclination ($i$), the longitude of ascending node ($\Omega$),
the longitude at reference epoch 2010.0 ($\lambda_\mathrm{ref}$), and an RV jitter term added in quadrature with
the measurement uncertainties ($\sigma_\mathrm{jit}$).
Other nuisance parameters such as parallax, barycentric proper motion,
and instrument-specific RV offsets are analytically marginalized out during the fits
for computational efficiency.
Priors in the fit are chosen to be broad or uninformative to avoid significantly influencing the posteriors.  
For the host mass we adopt a 
normal distribution centered at 1.02~\Msun \ based on the constraint from \citet{Luck:2017jd}
with a standard deviation of 0.10~\Msun, which is twice the uncertainty from Luck et al.
to mitigate potential systematic errors in the isochrones used in that analysis.
Log-uniform priors are adopted for the companion mass, semi-major axis, and RV jitter.
The remaining priors are uninformative: $\sin i$ is used for inclination, and 
uniform distributions are chosen for 
$\sqrt{e} \sin \omega$, $\sqrt{e} \cos \omega$, $\Omega$, and $\lambda_\mathrm{ref}$.
Eccentricities are bounded by [0,1), which is enforced by setting the prior equal to 0 if
$(\sqrt{e}  \cos \omega)^2 + (\sqrt{e} \sin \omega)^2 \ge 1$.  

Results of the orbit fit are shown in Figure~\ref{fig:orbit} and Table~\ref{tab:hd47127b_orbit}.
The best-fitting orbit (as determined by the maximum likelihood) implies
a semi-major axis of 74$^{+15}_{-22}$~AU, a modest eccentricity of 0.20$^{+0.45}_{-0.12}$, an inclination of 62$^{+6}_{-10}$$\degr$,
and an orbital period of 610$^{+160}_{-270}$~yr,
although all parameter posteriors are generally very broad as a result of the small fractional orbital coverage
of the observations.  
The companion mass distribution is also broad and right skewed with a mode at 95~\Mjup
and a median value of 105~\Mjup.  
The 95.4\% credible interval is
68--177~\Mjup, which spans the stellar-substellar boundary at $\approx$78~\Mjup.
This immediately rules out the possibility that the companion is a white dwarf. 
The implications of this mass constraint are further discussed below.

\section{Discussion and Conclusions}{\label{sec:discussion}}

Compared to isolated brown dwarfs, a key advantage of brown dwarf companions is that
their ages can be calibrated to that of their host star with the reasonable
assumption that both components are coeval.  
With an age of 7--10~Gyr,
HD 47127 B represents a rare instance of an old brown dwarf
with an independent age constraint. 
At these ages, evolutionary models generally predict effective temperatures
of $\approx$1300--2000~K 
for masses at or just below the substellar boundary (e.g., \citealt{Saumon:2008im}).
This corresponds to spectral types spanning early-L to early-T,
with the division between older stars and brown dwarfs occurring near L4 (\citealt{Dupuy:2017ke}).
Brown dwarfs with even lower masses ($\lesssim$70~\Mjup) should have cooled to T and Y dwarfs by the
age of HD 47127.

We recovered HD 47127 B in a single filter but we can nevertheless broadly predict
its expected spectral type using empirical relations between brown dwarf absolute magnitudes and spectral types.  
The $H$-band contrast from our December 2020 observations is 13.14~$\pm$~0.15~mag.  
The 2MASS $H$-band magnitude of HD 47127 is 5.28~$\pm$~0.03~mag (\citealt{Cutri:2003tp}),
implying $H$=18.42~$\pm$~0.15~mag and $M_H$=16.29~$\pm$~0.15~mag for HD 47127 B.\footnote{Note that
no filter correction between $H$ (MKO) and $H$ (2MASS) is applied for HD 47127 because this term is
much smaller than the uncertainty in our contrast measurement.}
This corresponds to a spectral type of $\approx$T7--T7.5 using the
mean absolute magnitudes from \citet{Dupuy:2012bp}.
Using the less precise October 2017 contrast gives $M_H$=15.6~$\pm$~0.6~mag
and a spectral type of $\approx$T5--T8 (within 2$\sigma$).
We therefore expect HD 47127 B to be a late-T dwarf, as illustrated in the 
near-infrared color-magnitude diagram in Figure~\ref{fig:orbit}.
However, follow-up photometry and spectroscopy are needed to establish the 
spectral type and determine physical properties of HD 47127 B.

Compared to other benchmark T dwarf companions, the $H$-band 
absolute magnitude of HD 47127 B is fainter than 
HR 2562 B, HD 13724 B, HD 19467 B, and Gl 229 B (which span $M_H$ = 14.2--15.6~mag),
but brighter than Gl 758 B and HD 4113 C ($M_H$ = 16.7--18.2~mag).
In terms of system architecture, HD 47127 B is perhaps most similar to HD 19467 B (\citealt{Crepp:2014ce}): 
their host stars are both main sequence Solar analogs (with spectral types of G5 and G3 and masses
close to 1~\Msun), the companions have mid- to late-T spectral types, and they 
orbit between 50--80 AU.

The posterior mass distribution for HD 47127 B implies its mass is $>$67.7~\Mjup \ 
with 99\% confidence.  However, HD 47127 B should be a T dwarf from its faint absolute magnitude
and must therefore be below the hydrogen burning limit of $\approx$78~\Mjup \ 
(\citealt{Burrows:2001wq})\footnote{Note that the substellar boundary is a function of 
metallicity, helium fraction, and cloud opacity and can range from $\approx$73--84~\Mjup \ (see, e.g., \citealt{Fernandes:2019ii}).}.  
Assuming HD 47127 B is single and the only companion causing
the observed acceleration, 
the best estimate for its mass is  67--78~\Mjup \ based on its dynamical mass and evolutionary models.
Another possibility that could account for the unusually large maximum likelihood value of 103~\Mjup \
from our orbit fits is if HD 47127 B itself is a close unresolved binary, 
such as a pair of $\approx$50+50~\Mjup \ T dwarfs.
Evolutionary models can be used to assess whether this binary hypothesis is plausible given the
absolute magnitude and age of HD 47127 B.  Cond models from \citet{Baraffe:2003bj}
predict that a single 50~\Mjup \ brown dwarf should have $M_H$ = 17.1 mag at 8.5~Gyr and a range of
$M_H$=16.8--17.4 mag at 7--10~Gyr.  An equal-flux binary would have $M_H$ = 16.3 mag at 8.5~Gyr
and $M_H$=16.0--16.6 mag at 7--10~Gyr.  This is in good agreement with the observed absolute magnitude of HD 47127 B,
indicating that two equal-mass brown dwarfs could resolve the mass-absolute magnitude discrepancy.
(Other binary mass ratios are of course also possible.)
In this scenario HD 47127 B would resemble $\epsilon$ Indi Bab and Gl 417 BC, substellar
companions that have been directly resolved into binaries.
There are also hints that HD 4113 C may be an unresolved binary 
based on a discrepancy between the measured mass, 
spectral type, and age of the system (\citealt{Cheetham:2018ha}).

It is also possible that another unseen close-in companion could be present in this system.
The residuals of our McDonald RVs after subtracting the linear acceleration 
would imply a 3-$\sigma$ upper limit of 24.6 m s$^{-1}$ for a velocity semi-amplitude.
This corresponds to planet minimum masses of $m_p \sin i$ $<$ \{0.3, 0.6, 0.9, 1.2, 2.0, 2.8\} \Mjup \ at 
orbital separations of \{0.1, 0.5, 1, 2, 5, 10\} AU, assuming circular orbits.
Planets with minimum masses below that of Jupiter can be ruled out within $\approx$1.5~AU. 
More massive planets and brown dwarfs can also reside at longer orbital periods
if the observed radial and astrometric acceleration of HD 47127 is the superposition of multiple companions,
as is suspected for the HD 206893 system (\citealt{Grandjean:2019cv}).

HD 47127 B is poised to become an important addition to the short list of benchmark T dwarf
companions with orbits and mass measurements.  
Additional relative astrometry of the companion and RV monitoring of its host
will refine its eccentricity  and mass, both of which are weakly constrained with
our observations.  
HD 47127 B orbits at a moderately wide separation of $\approx$2$''$, which makes 
it amenable to a wealth of follow-up studies
including spin ($v \sin{i}$) measurements and atmospheric characterization.
Radial velocities of the companion can help constrain its orbit and mass,
and high-resolution NIR spectroscopy with instruments like the Keck Planet Imager and Characterizer (\citealt{Mawet:2016aa})
can be used to test the binary hypothesis
by resolving multiple sets of lines.
HD 47127 B will be an especially good target for imaging and spectroscopy with the \emph{James 
Webb Space Telescope} to study its physical properties such as 
temperature, luminosity, surface gravity, and composition.

\facility{Smith (Tull Spectrograph), Keck:II (NIRC2)}

\acknowledgments

We appreciate the referee's prompt and constructive comments which improved the quality of this paper.
We also thank Zhoujian Zhang and Michael Liu for helpful comments on an early version of this manuscript. 
We are grateful to Erik Brugamyer, Caroline Caldwell, Candace Gray, Kevin Gullikson, Bryce Hobbs, Marshall Johnson, Diane Paulson, Paul Robertson, Ivan Ramirez, Zili Shen, Andrew Vanderburg, and Rob Wittenmyer for contributing to the Tull observations of HD 47127 presented in this study.
This work has benefitted from \emph{The UltracoolSheet}, maintained by Will Best, Trent Dupuy, Michael Liu, Rob Siverd, and Zhoujian Zhang, and developed from compilations by Dupuy \& Liu (2012, ApJS, 201, 19), Dupuy \& Kraus (2013, Science, 341, 1492), Liu et al. (2016, ApJ, 833, 96), Best et al. (2018, ApJS, 234, 1), and Best et al. (2020b, AJ, in press).

This work was supported by a NASA Keck PI Data Award, administered by the NASA Exoplanet Science Institute. Data presented herein were obtained at the W. M. Keck Observatory from telescope time allocated to the National Aeronautics and Space Administration through the agency's scientific partnership with the California Institute of Technology and the University of California. The Observatory was made possible by the generous financial support of the W. M. Keck Foundation.
B.P.B. acknowledges support from the National Science Foundation grant AST-1909209 and NASA Exoplanet Research Program grant 20-XRP20$\_$2-0119.
The authors wish to recognize and acknowledge the very significant cultural role and reverence that the summit of Maunakea has always had within the indigenous Hawaiian community. We are most fortunate to have the opportunity to conduct observations from this mountain.


\begin{thebibliography}{}
\expandafter\ifx\csname natexlab\endcsname\relax\def\natexlab#1{#1}\fi

\bibitem[{Ackerman \& Marley(2001)}]{Ackerman:2001gk}
Ackerman, A.~S., \& Marley, M.~S. 2001, ApJ, 556, 872

\bibitem[{Adams {et~al.}(1935)Adams, Joy, Humason, \& Brayton}]{Adams:1935aa}
Adams, W.~S., Joy, A.~H., Humason, M.~L., \& Brayton, A.~M. 1935, ApJ, 81, 187

\bibitem[{Baraffe {et~al.}(2003)Baraffe, Chabrier, Barman, Allard, \&
  Hauschildt}]{Baraffe:2003bj}
Baraffe, I., Chabrier, G., Barman, T.~S., Allard, F., \& Hauschildt, P.~H.
  2003, 402, 701

\bibitem[{Baraffe {et~al.}(2009)Baraffe, Chabrier, \&
  Gallardo}]{Baraffe:2009jw}
Baraffe, I., Chabrier, G., \& Gallardo, J. 2009, ApJ, 702, L27

\bibitem[{Bowler(2016)}]{Bowler:2016jk}
Bowler, B.~P. 2016, PASP, 128, 102001

\bibitem[{Bowler {et~al.}(2020)Bowler, Blunt, \& Nielsen}]{Bowler:2020hk}
Bowler, B.~P., Blunt, S.~C., \& Nielsen, E.~L. 2020, AJ, 159, 63

\bibitem[{Bowler {et~al.}(2015)Bowler, Liu, Shkolnik, \&
  Tamura}]{Bowler:2015ja}
Bowler, B.~P., Liu, M.~C., Shkolnik, E.~L., \& Tamura, M. 2015, ApJS, 216, 7

\bibitem[{Bowler {et~al.}(2018)Bowler, Dupuy, Endl, Cochran, MacQueen, Fulton,
  Petigura, Howard, Hirsch, Kratter, Crepp, Biller, Johnson, \&
  Wittenmyer}]{Bowler:2018gy}
Bowler, B.~P., Dupuy, T.~J., Endl, M., {et~al.} 2018, 155, 159

\bibitem[{Bowler {et~al.}(2021)Bowler, Cochran, Endl, Franson, Brandt, Dupuy,
  MacQueen, Kratter, Mawet, \& Ruane}]{Bowler:2021gg}
Bowler, B.~P., Cochran, W.~D., Endl, M., {et~al.} 2021, 161, 0

\bibitem[{Brandt(2018)}]{Brandt:2018dja}
Brandt, T.~D. 2018, ApJS, 239, 31

\bibitem[{Brandt {et~al.}(2019)Brandt, Dupuy, \& Bowler}]{Brandt:2019ey}
Brandt, T.~D., Dupuy, T.~J., \& Bowler, B.~P. 2019, 158, 140

\bibitem[{Burgasser {et~al.}(2006)Burgasser, Geballe, Leggett, Kirkpatrick, \&
  Golimowski}]{Burgasser:2006cf}
Burgasser, A.~J., Geballe, T.~R., Leggett, S.~K., Kirkpatrick, J.~D., \&
  Golimowski, D.~A. 2006, ApJ, 637, 1067

\bibitem[{Burrows {et~al.}(2001)Burrows, Hubbard, Lunine, \&
  Liebert}]{Burrows:2001wq}
Burrows, A., Hubbard, W.~B., Lunine, J.~I., \& Liebert, J. 2001, 73, 719

\bibitem[{Cheetham {et~al.}(2018)Cheetham, S{\'e}gransan, Peretti, Delisle,
  Hagelberg, Beuzit, Forveille, Marmier, Udry, \& Wildi}]{Cheetham:2018ha}
Cheetham, A., S{\'e}gransan, D., Peretti, S., {et~al.} 2018, A{\&}A, 614, A16

\bibitem[{Cochran \& Hatzes(1993)}]{Cochran:1993va}
Cochran, W.~D., \& Hatzes, A.~P. 1993, in ASP Conf. Ser. 36, Planets Around
  Pulsars, ed. J. A. Phillips, S. E. Thorsett, {\&} S. R. Kulkarni (San
  Francisco, CA: ASP), 267

\bibitem[{Crepp {et~al.}(2014)Crepp, Johnson, Howard, Marcy, Brewer, Fischer,
  Wright, \& Isaacson}]{Crepp:2014ce}
Crepp, J.~R., Johnson, J.~A., Howard, A.~W., {et~al.} 2014, ApJ, 781, 29

\bibitem[{Currie {et~al.}(2020)Currie, Brandt, Kuzuhara, Chilcote, Guyon,
  Marois, Groff, Lozi, Vievard, Sahoo, Deo, Jovanovic, Martinache, Wagner,
  Dupuy, Wahl, Letawsky, Li, Zeng, Brandt, Michalik, Grady, Janson, Knapp,
  Kwon, Lawson, McElwain, Uyama, Wisniewski, \& Tamura}]{Currie:2020hq}
Currie, T., Brandt, T.~D., Kuzuhara, M., {et~al.} 2020, ApJL, 904, 0

\bibitem[{Cushing {et~al.}(2011)Cushing, Kirkpatrick, Gelino, Griffith,
  Skrutskie, Mainzer, Marsh, Beichman, Burgasser, Prato, Simcoe, Marley,
  Saumon, Freedman, Eisenhardt, \& Wright}]{Cushing:2011dk}
Cushing, M.~C., Kirkpatrick, J.~D., Gelino, C.~R., {et~al.} 2011, ApJ, 743, 50

\bibitem[{Cutri {et~al.}(2003)Cutri, Skrutskie, Van~Dyk, Beichman, Carpenter,
  Chester, Cambresy, Evans, Fowler, Gizis, Howard, Huchra, Jarrett, Kopan,
  Kirkpatrick, Light, Marsh, McCallon, Schneider, Stiening, Sykes, Weinberg,
  Wheaton, Wheelock, \& Zacarias}]{Cutri:2003tp}
Cutri, R.~M., Skrutskie, M.~F., Van~Dyk, S., {et~al.} 2003, The 2MASS All-Sky
  Catalog of Point Sources, University of Massachusetts and Infrared Processing
  and Analysis Center; IPAC/California Institute of Technology

\bibitem[{Dupuy \& Liu(2012)}]{Dupuy:2012bp}
Dupuy, T.~J., \& Liu, M.~C. 2012, ApJS, 201, 19

\bibitem[{Dupuy \& Liu(2017)}]{Dupuy:2017ke}
---. 2017, ApJS, 231, 15

\bibitem[{Dupuy {et~al.}(2009)Dupuy, Liu, \& Ireland}]{Dupuy:2009wd}
Dupuy, T.~J., Liu, M.~C., \& Ireland, M.~J. 2009, ApJ, astro-ph.SR

\bibitem[{Dupuy {et~al.}(2014)Dupuy, Liu, \& Ireland}]{Dupuy:2014iz}
---. 2014, 790, 133

\bibitem[{Endl {et~al.}(2000)Endl, K{\"u}rster, \& Els}]{Endl:2000ui}
Endl, M., K{\"u}rster, M., \& Els, S. 2000, 362, 585

\bibitem[{Fernandes {et~al.}(2019)Fernandes, Van~Grootel, Salmon, Aringer,
  Burgasser, Scuflaire, Brassard, \& Fontaine}]{Fernandes:2019ii}
Fernandes, C.~S., Van~Grootel, V., Salmon, S. J. A.~J., {et~al.} 2019, ApJ,
  879, 0

\bibitem[{Fischer {et~al.}(2014)Fischer, Marcy, \& Spronck}]{Fischer:2014ew}
Fischer, D.~A., Marcy, G.~W., \& Spronck, J. F.~P. 2014, ApJS, 210, 5

\bibitem[{{Gaia Collaboration} {et~al.}(2020){Gaia Collaboration}, Brown,
  Vallenari, Prusti, \& de~Bruijne}]{GaiaCollaboration:2020ev}
{Gaia Collaboration}, Brown, A. G.~A., Vallenari, A., Prusti, T., \&
  de~Bruijne, J. H.~J. 2020, A{\&}A, 1

\bibitem[{Glebocki \& Gnacinski(2005)}]{Glebocki:2005aa}
Glebocki, R., \& Gnacinski, P. 2005, VizieR On-line Data Catalog, 3244

\bibitem[{Grandjean {et~al.}(2019)Grandjean, Lagrange, Beust, Rodet, Milli,
  Rubini, Babusiaux, Meunier, Delorme, Aigrain, Zicher, Bonnefoy, Biller,
  Baudino, Bonavita, Boccaletti, Cheetham, Girard, Hagelberg, Janson, Lannier,
  Lazzoni, Ligi, Maire, Mesa, Perrot, Rouan, \& Zurlo}]{Grandjean:2019cv}
Grandjean, A., Lagrange, A.-M., Beust, H., {et~al.} 2019, 627, L9

\bibitem[{Helling {et~al.}(2008)Helling, Ackerman, Allard, Dehn, Hauschildt,
  Homeier, Lodders, Marley, Rietmeijer, Tsuji, \& Woitke}]{Helling:2008gs}
Helling, C., Ackerman, A., Allard, F., {et~al.} 2008, MNRAS, 391, 1854

\bibitem[{Isaacson \& Fischer(2010)}]{Isaacson:2010gka}
Isaacson, H., \& Fischer, D. 2010, ApJ, 725, 875

\bibitem[{Kirkpatrick {et~al.}(1999)Kirkpatrick, Reid, Liebert, Cutri, Nelson,
  Beichman, Dahn, Monet, Gizis, \& Skrutskie}]{Kirkpatrick:1999ev}
Kirkpatrick, J.~D., Reid, I.~N., Liebert, J., {et~al.} 1999, ApJ, 519, 802

\bibitem[{Konopacky {et~al.}(2010)Konopacky, Ghez, Barman, Rice, Bailey, White,
  Mclean, \& Duch{\^e}ne}]{Konopacky:2010kr}
Konopacky, Q.~M., Ghez, A.~M., Barman, T.~S., {et~al.} 2010, ApJ, 711, 1087

\bibitem[{Kumar(1963)}]{Kumar:1963ht}
Kumar, S.~S. 1963, 137, 1121

\bibitem[{Lafreni{\`e}re {et~al.}(2007)Lafreni{\`e}re, Marois, Doyon, Nadeau,
  \& Artigau}]{Lafreniere:2007bg}
Lafreni{\`e}re, D., Marois, C., Doyon, R., Nadeau, D., \& Artigau, {\'E}. 2007,
  ApJ, 660, 770

\bibitem[{Liu(2004)}]{Liu:2004kk}
Liu, M.~C. 2004, Science, 305, 1442

\bibitem[{Liu {et~al.}(2008)Liu, Dupuy, \& Ireland}]{Liu:2008ib}
Liu, M.~C., Dupuy, T.~J., \& Ireland, M.~J. 2008, ApJ, 689, 436

\bibitem[{Luck(2017)}]{Luck:2017jd}
Luck, R.~E. 2017, 153, 21

\bibitem[{Maire {et~al.}(2020)Maire, Molaverdikhani, Desidera, Trifonov,
  Molli{\`e}re, D'Orazi, Frankel, Baudino, Messina, M{\"u}ller, Charnay,
  Cheetham, Delorme, Ligi, Bonnefoy, Brandner, Mesa, Cantalloube, Galicher,
  Henning, Biller, Hagelberg, Lagrange, Lavie, Rickman, S{\'e}gransan, Udry,
  Chauvin, Gratton, Langlois, Vigan, Meyer, Beuzit, Bhowmik, Boccaletti,
  Lazzoni, Perrot, Schmidt, Zurlo, Gluck, Pragt, Ramos, Roelfsema, Roux, \&
  Sauvage}]{Maire:2020iu}
Maire, A.~L., Molaverdikhani, K., Desidera, S., {et~al.} 2020, A{\&}A, 639, A47

\bibitem[{Mamajek \& Hillenbrand(2008)}]{Mamajek:2008jz}
Mamajek, E.~E., \& Hillenbrand, L.~A. 2008, ApJ, 687, 1264

\bibitem[{Marleau \& Cumming(2014)}]{Marleau:2014bh}
Marleau, G.~D., \& Cumming, A. 2014, MNRAS, 437, 1378

\bibitem[{Marley {et~al.}(2007)Marley, Fortney, Hubickyj, Bodenheimer, \&
  Lissauer}]{Marley:2007bf}
Marley, M.~S., Fortney, J.~J., Hubickyj, O., Bodenheimer, P., \& Lissauer,
  J.~J. 2007, ApJ, 655, 541

\bibitem[{Marois {et~al.}(2006)Marois, Lafreni{\`e}re, Doyon, Macintosh, \&
  Nadeau}]{Marois:2006df}
Marois, C., Lafreni{\`e}re, D., Doyon, R., Macintosh, B., \& Nadeau, D. 2006,
  ApJ, 641, 556

\bibitem[{Mawet {et~al.}(2016)Mawet, Wizinowich, Dekany, Chun, Hall, Cetre,
  Guyon, Wallace, Bowler, Liu, Ruane, Serabyn, Bartos, Wang, Vasisht,
  Fitzgerald, Skemer, Ireland, Fucik, Fortney, Crossfield, Hu, \&
  Benneke}]{Mawet:2016aa}
Mawet, D., Wizinowich, P., Dekany, R., {et~al.} 2016, in Proceedings of the
  SPIE, California Institute of Technology (United States), 99090D

\bibitem[{Morley {et~al.}(2012)Morley, Fortney, Marley, Visscher, Saumon, \&
  Leggett}]{Morley:2012io}
Morley, C.~V., Fortney, J.~J., Marley, M.~S., {et~al.} 2012, 756, 172

\bibitem[{Nelder \& Mead(1965)}]{Nelder:1965tk}
Nelder, J.~A., \& Mead, R. 1965, The Computer Journal, 7, 308

\bibitem[{Nordstrom {et~al.}(2004)Nordstrom, Mayor, Andersen, Holmberg, Pont,
  Jorgensen, Olsen, Udry, \& Mowlavi}]{Nordstrom:2004ci}
Nordstrom, B., Mayor, M., Andersen, J., {et~al.} 2004, 418, 989

\bibitem[{Paulson {et~al.}(2002)Paulson, Saar, Cochran, \&
  Hatzes}]{Paulson:2002tn}
Paulson, D.~B., Saar, S.~H., Cochran, W.~D., \& Hatzes, A.~P. 2002, 124, 572

\bibitem[{Rickman {et~al.}(2020)Rickman, S{\'e}gransan, Hagelberg, Beuzit,
  Cheetham, Delisle, Forveille, \& Udry}]{Rickman:2020aa}
Rickman, E.~L., S{\'e}gransan, D., Hagelberg, J., {et~al.} 2020, A{\&}A, 635,
  A203

\bibitem[{Saikia {et~al.}(2018)Saikia, Marvin, Jeffers, Reiners, Cameron,
  Marsden, Petit, Warnecke, \& Yadav}]{Saikia:2018dh}
Saikia, S.~B., Marvin, C.~J., Jeffers, S.~V., {et~al.} 2018, 616, A108

\bibitem[{Saumon \& Marley(2008)}]{Saumon:2008im}
Saumon, D., \& Marley, M.~S. 2008, ApJ, 689, 1327

\bibitem[{Spiegel \& Burrows(2012)}]{Spiegel:2012ea}
Spiegel, D.~S., \& Burrows, A. 2012, ApJ, 745, 174

\bibitem[{Stanford-Moore {et~al.}(2020)Stanford-Moore, Nielsen, De~Rosa,
  Macintosh, \& Czekala}]{StanfordMoore:2020kw}
Stanford-Moore, S.~A., Nielsen, E.~L., De~Rosa, R.~J., Macintosh, B., \&
  Czekala, I. 2020, ApJ, 898, 0

\bibitem[{Tull {et~al.}(1995)Tull, MacQueen, \& Sneden}]{Tull:1995tn}
Tull, R.~G., MacQueen, P.~J., \& Sneden, C. 1995, PASP, 107, 251

\bibitem[{Valenti \& Fischer(2005)}]{Valenti:2005fz}
Valenti, J.~A., \& Fischer, D.~A. 2005, ApJS, 159, 141

\bibitem[{Wright {et~al.}(2004)Wright, Marcy, Butler, \& Vogt}]{Wright:2004eb}
Wright, J.~T., Marcy, G.~W., Butler, R.~P., \& Vogt, S.~S. 2004, ApJS, 152, 261

\end{thebibliography}

\end{document}